\documentclass[citeautoscript,floatfix,longbibliography,bibnotes]{revtex4-2}
\usepackage{times,bm,amsmath,amsfonts,amsthm,amssymb,amsbsy,braket,hyperref,graphicx}
\usepackage[utf8]{inputenc}
\usepackage[T1]{fontenc}
\usepackage[capitalize]{cleveref}
\usepackage{microtype}
\hypersetup{colorlinks=true,urlcolor=blue,linkcolor=blue,citecolor=blue,pdfauthor={Pasquale Marra},pdftitle={Gauge fields induced by curved spacetime}}
\usepackage{physics}

\newcommand{\ii}{\mathrm{i}}
\DeclareMathOperator{\diag}{diag}

\begin{document}
\title{
Gauge fields induced by curved spacetime
}
\author{Pasquale Marra}
\email{pasquale.marra@keio.jp}
\affiliation{Department of Engineering and Applied Sciences, Sophia University, 7-1 Kioi-cho, Chiyoda-ku, Tokyo 102-8554, Japan}
\affiliation{Department of Physics, and Research and Education Center for Natural Sciences, Keio University, 4-1-1 Hiyoshi, Yokohama, Kanagawa, 223-8521, Japan}
\affiliation{Graduate School of Informatics, Nagoya University, Furo-cho, Chikusa-Ku, Nagoya, 464-8601, Japan}
\begin{abstract}
I found an extended duality (triality) between Dirac fermions in periodic spacetime metrics, nonrelativistic fermions in gauge fields (e.g., Harper-Hofstadter model), and in periodic scalar fields on a lattice (e.g., Aubry-André model).
This indicates an unexpected equivalence between spacetime metrics, gauge fields, and scalar fields on the lattice, which are understood as different physical representations of the same mathematical object, the quantum group $\mathcal{U}_q(\mathfrak{sl}_2)$.
This quantum group is generated by the exponentiation of two canonical conjugate operators, namely a linear combination of position and momentum (periodic spacetime metrics), the two components of the gauge-invariant momentum (gauge fields), and position and momentum (periodic scalar fields).
Hence, on a lattice, Dirac fermions in a periodic spacetime metric are equivalent to nonrelativistic fermions in a periodic scalar field after a proper canonical transformation.
The three lattice Hamiltonians (periodic spacetime metric, Harper-Hofstadter, and Aubry-André) share the same properties, namely fractal phase diagrams, self-similarity, $S$-duality, topological invariants, flat bands, and topologically quantized current in the incommensurate regimes.
In essence, this work unveils an unexpected link between gravity and gauge fields, opens new avenues for studying analog gravity, e.g., the Unruh effect and universe expansions/contractions, suggests the existence of an $S$-duality of spacetime curvatures, and hints at novel pathways to quantized gravity theories.
\end{abstract}
\date{\today}
\maketitle

\section{Introduction}

Gravity remains the lone outlier among the fundamental forces, defying all efforts to fit it into a coherent quantum theory --- a challenge that continues to puzzle physicists to this day.
However, in quantum field theory on curved spacetime~\cite{hollands_quantum_2015}, one usually ignores these issues and describes, e.g., fermions acting in the presence of gauge fields and within the backdrop of a spacetime metric.
Yet, even in this approach, gravity seems to have a special role, serving as the \emph{deus ex machina} that bends or reshapes the stage where gauge fields and fermions perform.
Hence, from the viewpoint of a fermion, a gauge field, a scalar field, and a spacetime metric are very different objects.

In the light of the above, the result obtained here is surprising:
On a discrete lattice~\cite{susskind_lattice_1977}, a periodic spacetime metric is equivalent to a gauge field and to a periodic scalar field.
More precisely, I found that the Hamiltonian of a Dirac fermion in a curved spacetime~\cite{mcvittie_diracs_1932,mann_semiclassical_1991} described by a periodic spacetime metric is equivalent to the Hamiltonian of a nonrelativistic fermion in a gauge field (Harper-Hofstadter model~\cite{harper_single_1955,hofstadter_energy_1976}) or in a periodic scalar field (Aubry-André model~\cite{aubry_analyticity_1980}) when regularized on the lattice.
Although the duality between nonrelativistic lattice fermions in gauge fields (Harper-Hofstadter) and periodic scalar fields (Aubry-André model) is well known~\cite{dominguez-castro_the-aubryandre_2019,marra_hofstadter-toda_2024}, the connection with Dirac fermions in curved periodic spacetime on a lattice is unexpected:
In essence, these three different models are all different physical representations of the same quantum group $\mathcal{U}_q(\mathfrak{sl}_2)$, generated by the exponentiation of two canonical conjugate operators.
These operators are the two components of the gauge-invariant momentum (Harper-Hofstadter model), position and momentum (Aubry-André model), and a linear combination of position and momentum (periodic spacetime metrics).
Hence, on a lattice, Dirac fermions in a periodic spacetime metric are obtained via a canonical transformation of nonrelativistic fermions in a periodic scalar field.
This triality (i.e., nontrivial equivalence between three different models) mandates that all three models share the same properties, e.g., fractal phase diagrams, self-similarity, topological invariants, flat bands, and topologically quantized current in the incommensurate regimes, as I will shortly illustrate.
Hamiltonians describing Dirac fermions in a curved spacetime regularized on a lattice are often considered in condensed matter~\cite{cortijo_a-cosmological_2007,juan_charge_2007,vozmediano_gauge_2008,de-juan_dislocations_2010,cortijo_geometrical_2012,iorio_quantum_2014,castro_symmetry_2018}.
These results may open new avenues for studying analog gravity, in particular the Unruh effect and universe expansions/contractions, and hint at novel ways to look at quantized gravity theories.

\section{Gauge fields and periodic scalar fields on a lattice}

Nonrelativistic fermions on a 2D discrete lattice in a gauge field are described by the so-called Harper-Hofstadter Hamiltonian.
To derive this Hamiltonian, one can start from charged fermions in a gauge field and then treat the discrete 2D lattice (described by a periodic potential) as a perturbation on the Landau levels (as in Ref.~\cite{thouless_quantized_1982}).
Conversely, one can start with a so-called "tight-binding" model describing charged fermions in a periodic potential and treating the gauge field as a perturbation (as in Refs.~\cite{harper_single_1955,hofstadter_energy_1976}).
These two approaches lead to the same result, namely
\begin{equation}\label{eq:HH2D}
\mathcal{H}_{\text{HH}}=
2 \cos{\hat{\pi}_x}+2\cos{\hat{\pi}_y},
\end{equation}
where the two terms are the contributions of the gauge-invariant momentum $\hat{\boldsymbol{\pi}}=\hat{\mathbf{p}}-\hat{\mathbf{A}}$ regularized on the lattice, where $\hat{\mathbf{A}}=(-\omega_y \hat{y},\omega_x \hat{x})$ of a uniform field in Coulomb gauge.
The two components of the gauge-invariant momentum are canonical conjugate operators $[\hat{\pi}_x,\hat{\pi}_y]=\ii\omega$ with $\omega=\omega_x+\omega_y$ equal to the flux of the field per unit cell.
In the continuum limit at small momenta, \cref{eq:HH2D} yields the Hamiltonian of charged particles in a uniform gauge field $\mathcal{H}_{\text{HH}} \sim (p_x+\omega_y y)^2 + (p_y-\omega_x x)^2$ (up to a constant value), which in the Landau gauge is equivalent to the Hamiltonian of the quantum harmonic oscillator $\mathcal{H}_{\text{HH}} \sim p_x^2 + \omega_x^2(x-k_y/\omega_x)^2$ where the minimum of the potential depends on the eigenvalue of the momentum $k_y$.
This is the celebrated Hamiltonian of the Landau levels that describes the physics of the integer quantum Hall effect of 2D electron systems, which exhibit topologically protected edge modes and quantized Hall conductance proportional to the topological Chern number (also known as the TKKN invariant introduced by Thouless, Kohomoto, Nightingale, and den Nijs~\cite{thouless_quantized_1982}).

On the other hand, nonrelativistic fermions on a 1D discrete lattice in a periodic scalar field are described by the so-called Aubry-André model~\cite{aubry_analyticity_1980}, which can be written as
\begin{equation}\label{eq:AA}
\mathcal{H}_{\text{AA}}=
2 \cos{\left( \omega\hat{x} + \phi \right)}+2\cos{\hat{p}},
\end{equation}
where the first term is the contribution of the scalar field $\propto\cos{\left( \omega{x} + \phi \right)}$, and the second term is the particle momentum regularized on the lattice.
In the continuum limit at small momenta and small frequencies $\omega$, \cref{eq:AA} yields $\mathcal{H}_{\text{AA}} \sim p^2 + \omega^2(x+\phi/\omega)^2$ (up to a constant value), which is again the quantum harmonic oscillator, now with the minimum of the potential depending on the phase shift $\phi$.
The Hamiltonian in \cref{eq:AA} can be generalized to $2 J\cos{\left( \omega\hat{x} + \phi \right)}+2K\cos{\hat{p}}$.
For almost every $\omega/2\pi,\phi/2\pi\in\mathbb{R}-\mathbb{Q}$ this Hamiltonian has exponentially localized eigenmodes in position space for $|J|>|K|$ and delocalized modes for $|J|<|K|$ with a phase transition, known as the Anderson localization transition, between the two regimes at $J=\pm K$~\cite{jitomirskaya_metal-insulator_1999}.
Hence, the Hamiltonian in \cref{eq:AA} coincides with the Aubry-André Hamiltonian sitting right at the phase boundary between the localized and delocalized phases.

The duality between $\mathcal{H}_{\text{HH}}$ and $\mathcal{H}_{\text{AA}}$ is evident: both Hamiltonians can be written as
\begin{equation}\label{eq:HXY}
\mathcal{H}_{XY}=
2 \cos{\hat{X}}+2\cos{\hat{Y}},
\end{equation}
where $\hat{X},\hat{Y}$ are the canonical conjugate operators $[\hat{X},\hat{Y}]=\ii\omega$.
Thus, taking $\hat{X}=\hat{p}_x$ and $\hat{Y}=\hat{p}_y$ yields the Harper-Hofstadter model in \cref{eq:HH2D};
taking $\hat{X}=\omega\hat{x}+\phi$ and $\hat{Y}=\hat{p}$ yields the Aubry-André model in \cref{eq:AA}.
The quantity $\omega=-\ii[\hat{X},\hat{Y}]$ describes the flux of the gauge field per unit cell in the Harper-Hofstadter model and the wavenumber of the periodic scalar field in the Aubry-André model.
The phase shift $\phi$ of the periodic scalar field in the Aubry-André model plays the same role as one of the components of the momentum in the Harper-Hofstadter model and, for this reason, can be regarded as a "synthetic" dimension~\cite{ozawa_topological_2019}.
In the language of quantum geometry~\cite{wiegmann_bethe-ansatz_1994,wiegmann_quantum_1994,faddeev_generalized_1995,kharchev_unitary_2002,hatsuda_hofstadters_2016,ikeda_hofstadters_2018,marra_hofstadter-toda_2024}, the Hamiltonian \cref{eq:HXY} is $\mathcal{H}_{XY}=\hat T_X+\hat T_X^\dag+\hat T_Y+\hat T_Y^\dag$ where $\hat T_X=e^{\ii \hat X}$ and $\hat T_Y=e^{\ii \hat Y}$ are the so-called magnetic translation operators, satisfying the algebraic equations $\hat T_X \hat T_Y=e^{\ii\omega}\hat T_Y \hat T_X$, which are a quantum "deformation" of the usual algebra of spatial translations, which is commutative $\hat T_X \hat T_Y=\hat T_Y \hat T_X$.
These operators $\hat T_X$ and $\hat T_Y$ coincide with two of the generators of the quantum group $\mathcal{U}_q(\mathfrak{sl}_2)$ obtained as a $q$-deformation of the enveloping algebra $U(\mathfrak{sl}_2)$ of the Lie algebra $\mathfrak{sl}_2$, with $q^2=e^{\ii\omega}$.

However, the story does not necessarily end here:
In principle, one can consider any set of conjugate variables $\hat{X},\hat{Y}$ in \cref{eq:HXY}.
An interesting choice is $\hat{X}=\frac12(\omega\hat{x}+\phi)-\hat{p}$ and $\hat{Y}=\frac12(\omega\hat{x}+\phi)+\hat{p}$ giving again $[\hat{X},\hat{Y}]=\ii\omega$.
Using the standard trigonometric sum-to-product identity, one gets
\begin{equation}\label{eq:G}
\mathcal{H}_\text{CS}=
4
\cos{\left(\frac12(\omega\hat{x}+\phi)\right)}
\cos{\hat{p}},
\end{equation}
where the subscript stands for \emph{curved spacetime}, as clarified hereafter.
In the continuum limit at small momenta and small frequencies, \cref{eq:G} yields $\mathcal{H}_{\text{CS}} \sim 2p^2 + \frac12\omega^2(x+\phi/\omega)^2$ (up to a constant value), which clearly describes a flat curvature geometry.
Regularizing \cref{eq:G} on the lattice
\begin{equation}\label{eq:Gposition}
\mathcal{H}_\text{CS}=
\sum\limits_n
2
\cos\left(\frac12(\omega n+\phi)\right)
\Big(
\ket{n+1}\bra{n}
+
\ket{n-1}\bra{n}
\Big)
.
\end{equation}
Note that this Hamiltonian coincides with the Su-Schrieffer-Bardeen model for polyacetylene chains~\cite{su_solitons_1979} for $\omega=2\pi$, and hence is a generalization of that model to arbitrary spatial frequencies $\omega\neq2\pi$.

Hence, not only the Hamiltonians $\mathcal{H}_{\text{HH}}$ and $\mathcal{H}_{\text{AA}}$ are dual, but they are also dual to the Hamiltonian $\mathcal{H}_{\text{CS}}$ in \cref{eq:G}, since all three Hamiltonians are formally expressed as in \cref{eq:HXY}.
In particular the Hamiltonian $\mathcal{H}_{\text{AA}}$ is related to the Hamiltonian in \cref{eq:G} by a canonical transformation $\hat{X}\to\frac12\hat{X}-\hat{Y}$, $\hat{Y}\to\frac12\hat{X}+\hat{Y}$.
Hence, the models $\mathcal{H}_{\text{HH}}$, $\mathcal{H}_{\text{AA}}$, and $\mathcal{H}_\text{CS}$ are dual one to the other:
I refer to this property as a triality.

To gain an intuition about the physical meaning of the Hamiltonian $\mathcal{H}_\text{CS}$, notice that in position basis, \cref{eq:Gposition} describes a fermion hopping back and forth with spatially modulated hopping amplitudes.
This spatial modulation intuitively suggests the presence of a warped or deformed spacetime metric.
This is because the covariant derivatives of a quantum field equation in curved spacetime typically correspond to space-dependent hopping amplitudes when regularized on the lattice.
Indeed, this intuition is correct, as I will show hereafter.

\section{Curved spacetime on a lattice}

Consider a massless Dirac fermion in 1+1D curved spacetime~\cite{mcvittie_diracs_1932,mann_semiclassical_1991}
\begin{equation}\label{eq:2DDirac}
 \left[\ii\gamma^a e_a{}^\mu \partial_\mu + \frac\ii2\gamma^a \frac1{\sqrt{-g}}\partial_\mu(\sqrt{-g}\,e_a{}^\mu)\right]\psi=0,
\end{equation}
where $\psi$ is a spinor, $\gamma^\mu$ the flat spacetime Dirac gamma matrices satisfying $\{\gamma^\mu, \gamma^\nu \}=2\eta^{\mu\nu}$ with $\eta_{\mu\nu}=\diag{(1,-1)}$ the Minkowski metric, $\sqrt{-g}$ the square root of the determinant of the metric, and the zweibein $g_{\mu\nu}= e^a{}_\mu e^b{}_\nu \eta_{ab}$, and $\eta_{ab}= e_a{}^\mu e_b{}^\nu g_{\mu\nu}$, with $g^{\mu\nu}=(g_{\mu\nu})^{-1}$.
In the Weyl representation $\gamma^0=\sigma_x$ and $\gamma^1=\ii\sigma_y$.
Consider the metric
\begin{equation}\label{eq:metric2}
\dd s^2=\alpha(x)^2 \dd t^2 - \dd x^2,
\end{equation}
which yields $g_{00}=\alpha(x)^2$ and $g_{11}=-1$, $e_0{}^0=\alpha(x)^{-1}$ and $e_1{}^1=1$, and $\sqrt{-g}=\alpha(x)$.
Separating time and space components
\begin{equation}
 \ii\partial_0 \psi=H\psi=
 -
{\ii\sqrt{\alpha(x)}}
\gamma_0 \gamma^1 \partial_1
 \left(
 \sqrt{\alpha(x)} \psi
 \right)
,
\label{eq:ContinuumHamiltonian2}
\end{equation}
which describes the time evolution of the spinor field, with $H$ the Hamiltonian density.
The corresponding Hamiltonian can be regularized on a discrete lattice using~\cite{marra_metric-induced_2024}
\begin{equation}
\partial_1\left(\sqrt{\alpha(x)}\psi\right)
\approx
\frac1{2}
\left(
\sqrt{\alpha_{n+1}}
\psi_{n+1}-
\sqrt{\alpha_{n-1}}
\psi_{n-1}
\right),
\end{equation}
where $\alpha_{n}=\alpha(n)$ is the discretized metric, which yields
\begin{equation}
\mathcal{H}=
\ii
\sum\limits_n
t_n
\psi_{n+1}^\dag
 \gamma_0\gamma^1
\psi_{n}
-
t_n
\psi_n^\dag
 \gamma_0\gamma^1
\psi_{n+1}
,
\end{equation}
where $t_n=\frac{1}{2}\sqrt{\alpha_n\alpha_{n+1}}$.
This Hamiltonian is hermitian.
The gauge transformation $\psi_n\to \ii^n\psi_n$ yields
\begin{equation}
\mathcal{H}=
\sum\limits_n
t_n
\left(
\psi_{n+1}^\dag
 \gamma_0\gamma^1
\psi_{n}
+
\psi_n^\dag
 \gamma_0\gamma^1
\psi_{n+1}
\right)
\label{eq:naturallyhermitianHamiltonian}
,
\end{equation}
which can be decomposed into two decoupled spin sectors in the Weyl representation $ \gamma_0\gamma^1 =-\sigma_z$.
Now, consider the metric
\begin{align}
{\alpha_{n}}=&
\frac1{C^{(-1)^{n}}}
\prod_{m=1}^{n-1}
\left[\cos^2\left(\frac12(\omega m+\phi)\right)\right]^{(-1)^{m+n+1}}
\nonumber\\
=&
\begin{cases}
C
\frac{
\cos^2\left(\frac12(2\omega+\phi)\right)
\cos^2\left(\frac12(4\omega+\phi)\right)
\cdots
\cos^2\left(\frac12((n-1)\omega+\phi)\right)
}
{
\cos^2\left(\frac12(1\omega+\phi)\right)
\cos^2\left(\frac12(3\omega+\phi)\right)
\cdots
\cos^2\left(\frac12((n-2)\omega+\phi)\right)
}
& \text{for odd } n,\\[4mm]
\frac1C
\frac{
\cos^2\left(\frac12(1\omega+\phi)\right)
\cos^2\left(\frac12(3\omega+\phi)\right)
\cdots
\cos^2\left(\frac12((n-1)\omega+\phi)\right)
}
{
\cos^2\left(\frac12(2\omega+\phi)\right)
\cos^2\left(\frac12(4\omega+\phi)\right)
\cdots
\cos^2\left(\frac12((n-2)\omega+\phi)\right)
}
& \text{for even } n,\\
\end{cases}
\label{eq:specialmetric}
\end{align}
for $n\ge1$ and with $C$ an arbitrary number defining the boundary value $\alpha_1=C$.
In this metric $2 t_n=\sqrt{{\alpha_n \alpha_{n+1}}}=|\cos{\left(\frac12(\omega n+\phi)\right)}|$, and therefore each of the two decoupled spin sectors of the Hamiltonian $\mathcal{H}$ in \cref{eq:naturallyhermitianHamiltonian} becomes equivalent to the Hamiltonian $\mathcal{H}_\text{CS}$ in \cref{eq:G} describing massless Dirac fermions in a periodic spacetime metric on a 1D lattice.

When the spatial frequency is commensurate with the lattice, i.e., when $\omega=2\pi p/q$ with $p,q$ coprimes, the hopping amplitudes are periodic.
Conversely, when the spatial frequency is incommensurate, the hopping amplitudes are quasiperiodic.
Note that the metric on the lattice $\alpha_n$ diverges for values of the phase $\omega m+\phi=\pm\pi$ on lattice sites where $m+n$ is even.
The values of the phase where the metric diverges form a set $\{\phi=\pm\pi-\omega m \mod 2\pi, m\in\mathbb{Z}\}\subset[0,2\pi]$.
In the commensurate case, the cardinality of the set is $q$ or $2q$, depending on whether $q$ is even or odd, i.e., there is a finite number of values ($q$ or $2q$) of the phase $\phi$ where the metric diverges.
In the incommensurate case, the cardinality of the set is $\aleph_0$, i.e., there is an infinite (and countable) number of values of the phase where the metric diverges.
Note that, even in cases where the metric diverges, the hopping amplitudes $t_n\propto\sqrt{\alpha_n \alpha_{n+1}}$ remain finite and well-defined, since in the limit $\phi\to\pm\pi-\omega m$, the quantity $\alpha_n \alpha_{n+1}$ converges on the whole lattice to a finite positive value or to zero, e.g., $t_m=0$.
Note also that $\alpha_n$ is not necessarily periodic, in contrast with the hopping amplitudes $t_n$ which are always periodic in the commensurate regime or quasiperiodic in the incommensurate regime.
In particular, one can show that $\alpha_n$ is periodic on the lattice with period $2q$ in the commensurate regime only for $q$ odd (see proof in Appendix A).
However, for the sake of simplicity, I refer to the metric as being periodic or quasiperiodic as long as the hopping amplitudes on the lattice are periodic or quasiperiodic.
For $p=1$, $C=1$, and large values of $q$, corresponding to the limit $\omega\approx0$, I empirically found that, by taking $\phi=0$ in \cref{eq:specialmetric} when $q$ is odd, and $\phi=\pi/q$ when $q$ is even, the metric can be approximated as
\begin{equation}
\alpha_n\approx
\left\vert\cos(\frac{n\pi}q - \frac{\pi}{2q}+\frac{\phi}2)\right\vert,
\label{eq:metriclimit}
\end{equation}
which becomes exact in the limit $q\to\infty$, with the difference between $\alpha_n$ and its approximate value scaling as $\propto 1/q$, as shown in Appendix B.
In this limit, the metric $\alpha_n$ is periodic with period equal to $q$, that is, $\alpha_{n+q}=\alpha_n$, regardless of the value of $q$.
Indeed, the metric is periodic also when $q$ is even (if $\phi=\pi/q$), which is not necessarily the case for even $q$ with arbitrary phases $\phi$ and arbitrary $p$.
Notice also that in this limit, the metric becomes linear at $n=q/2$, as also revealed in \cref{fig:example}(i).
Finally, the explicit form of the metric in \cref{eq:specialmetric} for small values of $p,q$ is given in Appendix C.

The lattice Hamiltonian $\mathcal{H}_\text{CS}$ describes fermions localized on a discrete set of points on a 1D line by the effect of, e.g., a scalar potential, and affected by the presence of a spacetime curvature in \cref{eq:metric2}.
Moreover, $\mathcal{H}_\text{CS}$ can be effectively reproduced in condensed matter by engineering a 1D lattice of fermions localized in ionic crystals or in optical lattices with variable hopping amplitudes.

\begin{figure}[t]
 \centering
		\includegraphics[width=.7\textwidth]
	{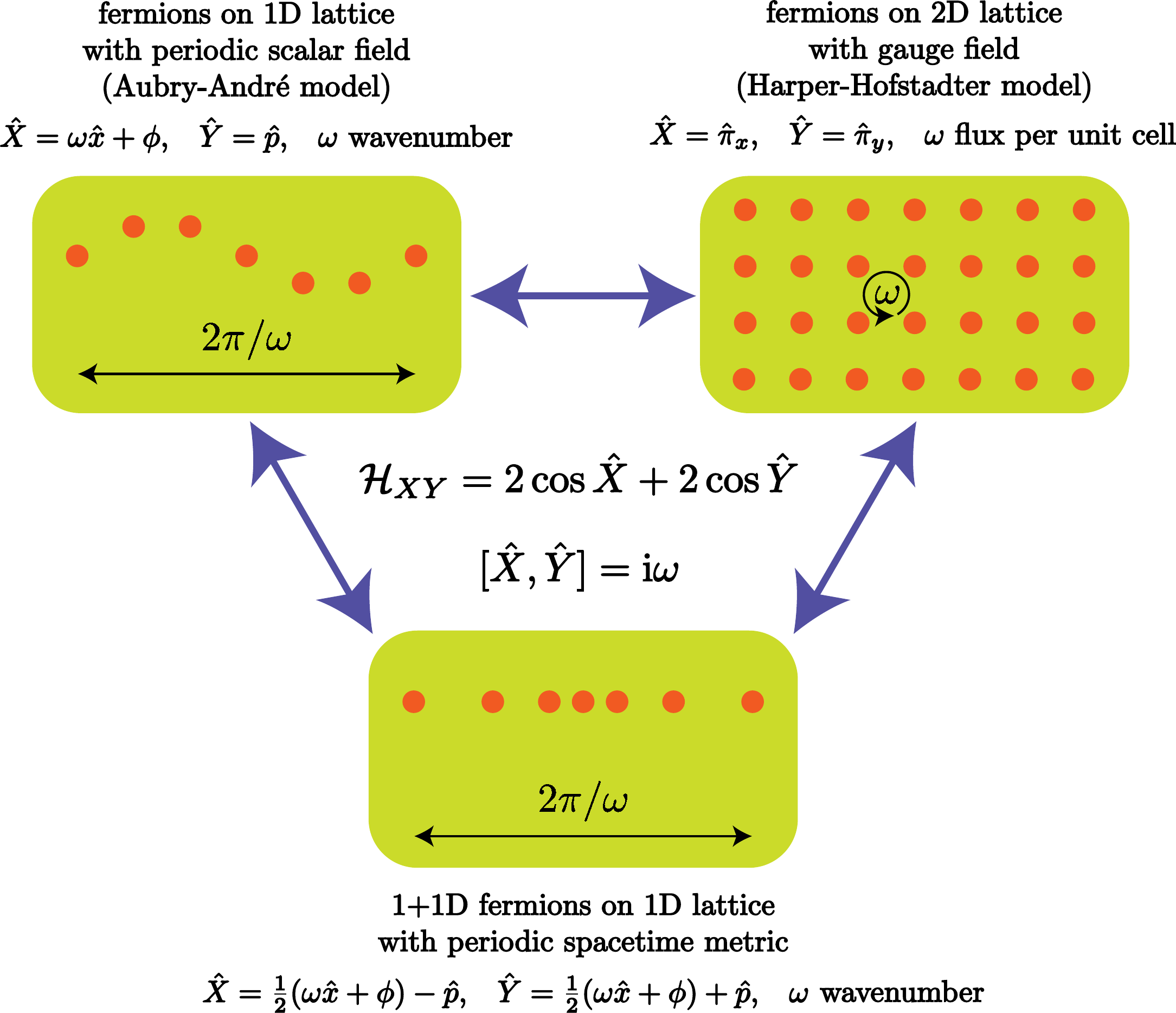}
	\caption{
Triality between gauge fields, periodic scalar fields, and curved spacetime metrics on finite lattices.
The Hamiltonian $\mathcal{H}_\text{AA}$ (left) of a nonrelativistic charged fermion on a 1D lattice in a periodic scalar field with frequency $\omega$ and phase shift $\phi$ (Aubry–André model) is equivalent under the transformation $(\omega \hat x, \hat p) \to (\hat{\pi}_x,\hat {\pi}_y)$ to the Hamiltonian $\mathcal{H}_\text{HH}$ (right) of a nonrelativistic charged fermion on a 2D square lattice in a gauge field with flux $\omega$ per unit cell (Harper-Hofstadter model).
These two Hamiltonians are equivalent under a canonical transformation to the Hamiltonian $\mathcal{H}_\text{CS}$ (bottom) of a massless relativistic fermion in a periodic spacetime metric with frequency $\omega$.
These three Hamiltonians $\mathcal{H}_\text{AA}$, $\mathcal{H}_\text{HH}$, and $\mathcal{H}_\text{CS}$ can all be written as $\mathcal{H}_{XY}=2\cos\hat X+2\cos\hat Y$ where $[\hat X,\hat Y]=\ii\omega$ and with the two canonical conjugate variables being respectively (left) $\hat X,\hat Y=\omega \hat x,\hat p$, (right) $\hat X,\hat Y=\hat{\pi}_x,\hat{\pi}_y$, and (bottom) $\hat X,\hat Y=\frac12\omega \hat x+\hat p,\frac12\omega \hat x-\hat p$.
}
\label{fig:triality}
\end{figure}

The lattice Hamiltonians $\mathcal{H}_{\text{HH}}$ and $\mathcal{H}_{\text{AA}}$ describing a massive nonrelativistic fermion respectively in a gauge field and in a periodic scalar field, together with the Hamiltonian $\mathcal{H}_\text{CS}$ describing a massless Dirac fermion in a periodic spacetime metric, are dual one to the other, being three different incarnations of the general Hamiltonian $\mathcal{H}_{XY}$ in \cref{eq:HXY}.
This triality is summarized in \cref{fig:triality}.
Note that the role of the quantity $\omega$ is played by the flux of the gauge field per unit cell, the spatial frequency of the periodic scalar field, and the spatial frequency of the spacetime metric, respectively, for $\mathcal{H}_{\text{HH}}$, $\mathcal{H}_{\text{AA}}$, and $\mathcal{H}_\text{CS}$.

\section{Topology, fractality, flat bands, and quantum geometry}

The triality mandates that every property valid for any of the three Hamiltonians is valid, \emph{mutatis mutandis}, for all the other two.
I will now go through some of these properties.

The topological properties of the models are encoded in a topological invariant~\cite{thouless_quantized_1982,thouless_quantization_1983,niu_quantum_1987}, the first Chern number, which is an integer defined as $c=\frac 1{2\pi}\sum_{n}^\text{N}\int_{T^2}\Omega_n(\boldsymbol \kappa) \mathrm d^2 \boldsymbol \kappa$, where $N$ is the number of occupied fermionic levels, $\Omega_n(\boldsymbol \kappa)=\nabla_{\boldsymbol\kappa}\mathcal{A}_n(\boldsymbol \kappa)$ the Berry curvature corresponding to the Berry connection
$\mathcal{A}_n(\boldsymbol \kappa)=\ii\bra{u_{n}(\boldsymbol \kappa)} \nabla_{\boldsymbol \kappa} \ket{u_{n}(\boldsymbol \kappa)}$ of the eigenvectors $\ket{u_{n}(\boldsymbol \kappa)}$ (see, e.g., Ref.~\cite{stanescu_introduction_2017_541}).
Here, the quantum numbers $\boldsymbol\kappa=(k_x,k_y)$ in the language of the Hamiltonian $\mathcal{H}_{\text{HH}}$, and $\boldsymbol\kappa=(\phi,k)$ in the language of the Hamiltonians $\mathcal{H}_{\text{AA}}$ and $\mathcal{H}_{\text{CS}}$, and the integral is over the whole space spanned by $\boldsymbol\kappa$, which is a torus $T^2$ (i.e., the Brillouin zone) due to the periodicity of the spectra with respect to $\kappa_x$ and $\kappa_y$.
Due to the bulk-boundary correspondence~\cite{hatsugai_chern_1993,ryu_topological_2002,teo_topological_2010}, a number $c$ of topologically protected edge modes appear at the edges of a topologically nontrivial phase $c\neq0$.

If the spatial frequency is commensurate with the lattice, the spectrum of the Hamiltonian $\mathcal{H}_\text{CS}$ in \cref{eq:G} has exactly $q$ energy levels spanning the phase $\phi$ and momentum $k$.
Each of the open gaps $j$ is labeled by a nonzero topological invariant $c\in\mathbb{Z}$ given by the Chern number~\cite{thouless_quantization_1983,thouless_quantized_1982} $|c|<q/2$ satisfying the diophantine equation~\cite{bellissard_the-noncommutative_1994,bellissard_the-noncommutative_2003,osadchy_hofstadter_2001} $p c = j \mod q$.
These nontrivial gaps exhibit $|c|$ topologically protected edge modes localized at the boundaries when considering a finite lattice of $N=mq$ sites with open boundary conditions, as it follows from the bulk-boundary correspondence~\cite{hatsugai_chern_1993}.

For incommensurate spatial frequencies $\omega$, i.e., when $\omega/2\pi$ is an irrational number $\in\mathbb{R}-\mathbb{Q}$, the energy spectrum is a Cantor set~\cite{avila_solving_2006}, and the dispersion of the Bloch bands becomes flat~\cite{kraus_topological_2012,harper_perturbative_2014,marra_topologically_2020}.
Hence, the periodic spacetime metric induces a fractal phase diagram spanned by the spatial frequency $\omega$ with infinitely many nontrivial gaps, as in the Hofstadter butterfly~\cite{hofstadter_energy_1976}.
For almost every $\omega/2\pi,\phi/2\pi\in\mathbb{R}-\mathbb{Q}$, this regime coincides with the phase boundary (Anderson localization transition) between the localized and delocalized phases in the Aubry-André model in \cref{eq:AA}~\cite{jitomirskaya_metal-insulator_1999}.

This phase diagram shows a remarkable self-similarity property~\cite{hatsuda_hofstadters_2016,ikeda_hofstadters_2018}:
It is invariant under the action of the duality transformation $p/q\to q/p,E\to\widetilde E$, with $\widetilde E$ given by some unknown function and with $p/q$ defined modulo 1 (or equivalently $\omega=2\pi p/q$ defined modulo $2\pi$).
This property has been related to the notion of modular double~\cite{faddeev_modular_2014,hatsuda_hofstadters_2016}, to the Langlands duality of quantum groups~\cite{ikeda_hofstadters_2018} and, by extension,
to the $S$-duality~\cite{hatsuda_hofstadters_2016,ikeda_hofstadters_2018}
(which generalized the symmetry under interchange of electric and magnetic fields in electromagnetism).
In the Harper-Hofstadter model, the two branches of the $S$-duality transformation $p/q\to q/p$ correspond to two opposite limits, obtained by considering the periodic potential as a perturbation on the Landau levels, or by considering the gauge field as a perturbation on the tight-binding model describing particles trapped in a periodic potential, as already noted in Ref.~\cite{thouless_quantized_1982}.
These two limits are obtained by exchanging the role of the electric field (periodic potential) and of the magnetic field (gauge field).
In the Aubry-André model and in the curved spacetime model in \cref{eq:G}, this duality transformation corresponds to switching the spatial frequency $\omega\to(2\pi)^2/\omega$ (modulo $2\pi$) of the periodic scalar field and of the periodic spacetime metric, respectively.
In the curved spacetime model, this $S$-duality transformation corresponds to switching the spatial frequency $\omega\to(2\pi)^2/\omega$ (modulo $2\pi$) of the curved spacetime metric.
Hence, the triality we unveiled here allows one to extend the notion of $S$-duality and Langlands duality to the duality between different periodic spacetimes, specifically, between spacetime curvatures with different spatial frequencies.
In particular, let us consider a simple example.
At small spatial frequencies (long wavelengths), in the limit $\omega=2\pi/q$ obtained by taking $p=1$ and large $q$, where the metric can be approximated as in \cref{eq:metriclimit}, the duality transformation yields $2\pi/q\to0$ (modulo $2\pi$).
In this case, the dual spacetime becomes the flat spacetime with $\omega=0$:
This implies the existence of an $S$-duality (and Langlands duality) between spacetimes with periodic curvatures and flat spacetimes.
The energy spectrum of the Hamiltonian is given by the roots of the characteristic polynomial expressed by the Chambers relation~\cite{chambers_linear-network_1965}
\begin{gather}
\det(\mathcal{H}_\text{CS}-E)=
\det(\mathcal{H}_{\text{AA}}-E)=
\nonumber\\
f_{p/q}(E)-2(-1)^q\left(\cos(q k)+\cos(q\phi)\right),
\label{eq:chambers}
\end{gather}
where with abuse of notation, $\mathcal{H}_\text{CS}$, $\mathcal{H}_\text{AA}$ denote here the matrices expressing the corresponding Hamiltonians in the basis of plane waves $e^{\ii k n}$ on the lattice, with the function $f_{p/q}(E)$ given by a polynomial in $E$ (see Ref.~\cite{marra_hofstadter-toda_2024}).
An analogous expression for $\det(\mathcal{H}_{\text{HH}}-E)$ is obtained by taking $k\to k_x$ and $\phi\to k_y$.
Remarkably $f_{p/q}(E)=f_{q/p}(\widetilde E)$ under the duality transformation $p/q\to q/p$~\cite{hatsuda_hofstadters_2016,ikeda_hofstadters_2018}.

In the Aubry-André model, nonzero Chern numbers correspond to the quantization of the charge transport and to the so-called topological charge pump~\cite{thouless_quantization_1983}.
The same can be said for the spacetime metric model Hamiltonian $\mathcal{H}_\text{CS}$:
During an adiabatic evolution of the phase $\phi=0$ to $2\pi$, the center of mass of $j$ particles on the lattice describes a trajectory that corresponds to a finite number of particles (i.e., amount of charge) transferred through the lattice, and this number is an integer equal to the Chern number $c$.
In the incommensurate case, moreover, not only the charge transferred is quantized, but also the corresponding current:
Precisely, the pumped charge becomes linear in time, and the instantaneous current (i.e., the time derivative of the charge pumped) becomes constant, being topologically quantized and equal to the Chern number~\cite{marra_topologically_2020}.
Hence, one obtains a quantization of the current induced by the spacetime metric on a lattice.

\section{Discussion}

\Cref{eq:G} is, in essence, a discrete tight-binding Hamiltonian with lattice-dependent hopping amplitudes.
As such, there are several physical systems that simulate this model, e.g., arrays of lattices of atoms deposited on a surface~\cite{drost_topological_2017,palacio-morales_atomic-scale_2019}, arrays of quantum dots~\cite{zhao_large-area_2021}, cold atoms in optical lattices~\cite{bloch_many-body_2008,boada_dirac_2011,minar_mimicking_2015,nakajima_topological_2016,lohse_a-thouless_2016,mula_casimir_2021,nakajima_competition_2021}, photonic crystals~\cite{ozawa_topological_2019}, superconducting quantum circuits~\cite{houck_on-chip_2012,anderson_engineering_2016,gu_microwave_2017}, topologically nontrivial stripes~\cite{marra_majorana_2024}, and exciton-polariton condensates in artificial lattices~\cite{valle-inclan-redondo_non-reciprocal_2024}.
Controlling the hopping amplitudes corresponds to controlling the distances and overlap amplitudes between contiguous localized states on the lattice.
Lattice models of curved spacetime provide ways to simulate gravitational effects in condensed matter, such as the Hawking/Unruh effect~\cite{rodriguez-laguna_synthetic_2017,sabsovich_hawking_2022,horner_chiral_2023,benhemou_probing_2023,maertens_hawking_2024} or AdS/CFT~\cite{yang_simulating_2020,deger_ads/cft_2023}
or to use geometric language to describe condensed matter phenomena~\cite{golan_probing_2018,castro_symmetry_2018}.

\begin{figure*}[t]
 \centering
	\includegraphics[width=1\textwidth]{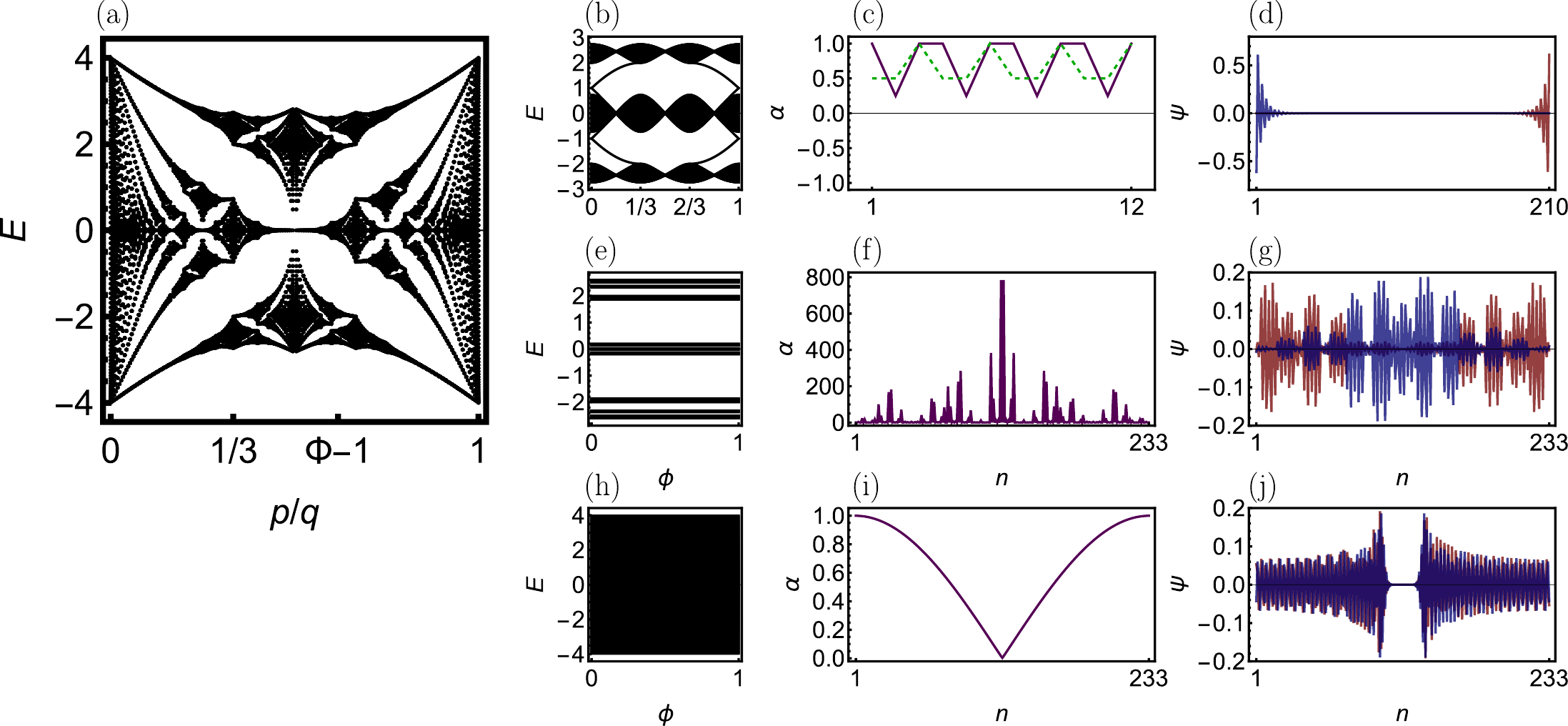}
	\caption{
Energy spectra of the Hamiltonian $\mathcal{H}_\text{CS}$ of a massless Dirac fermion in a periodic spacetime metric on a lattice of $N=210$ sites.
(a)
Energy spectra as a function of $p/q$ with periodic boundary conditions and $\phi=0$.
(b)
Energy spectra for $p/q=1/3$ as a function of the phase $\phi$ with open boundary conditions.
(c)
The metric $\alpha_n$ for $p/q=1/3$ and $\phi=0$ compared with $
\cos{\left(\frac12(\omega n+\phi)\right)}$ (dashed) as a function of the lattice site.
(d)
Wavefunctions of the two edge modes in the first band gap (from below) for $p/q=1/3$ and $\phi=0$.
(e)
Energy spectra for $\omega\approx2\pi(\Phi-1)$ (with $\Phi$ the golden ratio), as a function of the phase $\phi$ with periodic boundary conditions on a lattice of $N=233$ sites.
(f)
The metric $\alpha_n$ for $\omega\approx2\pi(\Phi-1)$ and $\phi=0$ as a function of the lattice site.
(g)
Wavefunctions of two bulk modes for $\omega\approx2\pi(\Phi-1)$ and $\phi=0$.
(h)
Energy spectra for small $\omega\approx0$, as a function of the phase $\phi$ with periodic boundary conditions on a lattice of $N=233$ sites.
The spectra show several discrete levels at a short distance, which corresponds to a continuum of energy levels in the limit $\omega\to0$.
(i)
The metric $\alpha_n$ for $\omega=2\pi/233\approx0$ (taking $p/q=1/233$) and $\phi=0$ as a function of the lattice site.
Notice that this metric becomes linear at the center $n=q/2$.
(j)
Wavefunctions of two bulk modes for $\omega\approx0$ and $\phi=0$.
}
\label{fig:example}
\end{figure*}

\Cref{fig:example}(a) shows the energy spectra of the Hamiltonian $\mathcal{H}_\text{CS}$ (Hofstadter butterfly) describing massless Dirac fermions in curved spacetime on a lattice as a function of the spatial frequency $\omega$ of the periodic spacetime metric.
Figures~\ref{fig:example}(b) to~\ref{fig:example}(d) show the energy spectrum and edge modes calculated with open boundary conditions for $p/q=1/3$ and the corresponding periodic spacetime metric in \cref{eq:specialmetric}.
Figures~\ref{fig:example}(e) to~\ref{fig:example}(g) show the energy spectrum and some of the bulk modes calculated with periodic boundary conditions and the corresponding quasiperiodic spacetime metric in \cref{eq:specialmetric} for a spatial frequency incommensurate with the lattice, specifically $\omega=2\pi(\Phi-1)$, where $\Phi$ is the golden ratio.
Note that the dispersion of the energy bands in the synthetic dimension $\phi$ is flattened, corresponding to energy modes delocalized on the entire lattice.
Note also that the spacetime metric is not periodic anymore and exhibits large amplitude oscillations.
Figures~\ref{fig:example}(h) to~\ref{fig:example}(j) show the energy spectrum and some of the bulk modes with the corresponding spacetime metric for a small spatial frequency $\omega=2\pi/N$, with $N$ the number of lattice sites.
Notice that this metric becomes linear at the center $n=N/2$ of the system.

Note that the metric in \cref{eq:metric2} yields the familiar Rindler metric if $\alpha(x)=rx$.
The Rindler metric is the metric of a reference frame in uniform acceleration in flat spacetime, and has a singularity at $x=0$, where the metric tensor has zero determinant and approximates the Schwarzschild metric near a black hole horizon.
This metric is approximately realized by \cref{eq:specialmetric} for small frequencies $\omega$ in finite patches of the spacetime in the case where the metric does not exhibit divergencies [see, e.g., \cref{fig:example}(i)].
This setup allows to simulate a black hole event horizon and in particular the Unruh effect, i.e., the emergence of a thermal distribution of excited states at zero-temperature induced by the presence of the event horizon~\cite{fulling_nonuniqueness_1973,unruh_notes_1976,davies_scalar_1975,unruh_experimental_1981,takagi_vacuum_1986}, with the resulting a nonzero effective temperature analogous of Hawking or Unruh radiation~\cite{unruh_notes_1976,unruh_experimental_1981}.
In this setup, the Unruh radiation is produced by a sudden quench corresponding to an abrupt change in the hopping term (i.e., the spacetime curvature).
Note that these phenomena are not experimentally accessible in high-energy experiments, since the Universe we live in is locally flat.
Moreover, by allowing a time-dependence of the spacetime curvature, the Hamiltonian $\mathcal{H}_\text{CS}$ allows to simulate dilating and contracting universes, which is again not directly observable in our universe timeline.

\section{Conclusions}

In conclusion, I extended the duality between the Harper-Hofstadter model and the Aubry-André model, describing lattice fermions in a gauge field and in a periodic scale field, to a triality between these two models and a model describing lattice Dirac fermions in a periodic spacetime metric.
This unveils an unexpected equivalence between spacetime metrics, gauge fields, and scalar fields on the lattice.
This triality is the equivalence of these three different models being different physical representations of the same quantum group $\mathcal{U}_q(\mathfrak{sl}_2)$.
This quantum group is generated by the exponentiation of two canonical conjugate operators, namely the two components of the gauge invariant momentum (Harper-Hofstadter model), position and momentum (Aubry-André model), and a linear combination of position and momentum (periodic spacetime metric).
Hence, on the lattice, a Dirac fermion in a periodic spacetime metric is equivalent to a nonrelativistic fermion in a periodic scalar field after a proper canonical transformation.
I thus derived several properties of Dirac fermions on a lattice in a periodic spacetime metric by applying the triality, namely, the fractality of the phase diagram, self-similarity properties, topological invariants, flat bands, and topological quantized current in the incommensurate regimes.

On a more fundamental level, the triality can lead to novel ways to look at quantum gravity theories.
Notably, the triality between gauge and spacetime curvature implies that the $S$-duality of the Harper-Hofstadter model, where the roles of the electric and magnetic fields are exchanged by taking $p/q\to q/p$, can be extended to a duality between spacetime curvatures with reciprocal wavelengths, again by taking $p/q\to q/p$.

We stress the fact that the triality unveiled here essentially requires the presence of a discrete lattice, while it becomes trivial in the continuum (where the spacetime curvature becomes flat).
For this reason, this triality may have implications in the context of loop quantum gravity and other quantum gravity theories that assume a discrete spacetime geometry at the Planck scale~\cite{rovelli_spin_1995}.

Finally, we notice that the spacetime curvature considered here is classical (not quantized) while the fermion fields are quantized.
Hence, one can extend the model considered here to a quantized gravity model such as that studied in the context of quantum gravitational lensing in Ref.~\cite{kaku_quantumness_2022,kaku_gravitational_2025} by promoting the curvature component to operators $\alpha(x)\to \alpha(\hat x)$.
Another approach is to consider entangled states induced by the superposition of two curvatures corresponding, e.g., to the Newtonian potential in a Schrödinger-Newton gravity model.
This leads to a condensed matter analog of quantum gravity, which is, in turn, equivalent to a system with flat curvature and a finite gauge field, due to the triality.
Hence, the triality opens the door to studying quantized gravity models via studying the better-understood gauge fields in quantum electrodynamics (QED).
This may hint at novel directions in pursuing a unified theory of quantum mechanics and general relativity.

\begin{acknowledgments}
I acknowledge the stimulating discussions with Antonino Flachi, Dongsheng Ge, and Youka Kaku.
This work is partially supported by the Japan Society for the Promotion of Science (JSPS) Grant-in-Aid for Early-Career Scientists Grants No.~23K13028 and No.~20K14375, and by the Japan Science and Technology Agency (JST) of the Ministry of Education, Culture, Sports, Science and Technology (MEXT), JST CREST Grant.~No.~JPMJCR19T2, Grant-in-Aid for Transformative Research Areas (A) KAKENHI Grant~No.~22H05111, and Grant-in-Aid for Transformative Research Areas (B) KAKENHI Grant~No.~24H00826.
\end{acknowledgments}

\appendix
\section{Analytical demonstration that the metric is periodic for commensurate spatial frequencies}

Here, I demonstrate that
\begin{equation}
{\alpha_{n}}=
C^{(-1)^{n}}
\prod_{m=1}^{n-1}
\left[\cos^2\left(\frac12(\omega m+\phi)\right)\right]^{(-1)^{m+n+1}},
\end{equation}
as defined in \cref{eq:specialmetric}, is a periodic function of the lattice index $n$ with period $2q$ when $\omega=2\pi p/q$ with $p,q$ coprimes and $q$ odd.
Assume $C=1$, and notice that for $n>2q$ one has
\begin{align}
{\alpha_{n+2q}}=&
\prod_{m=1}^{2q}
\left[\cos^2\left(\frac{p m \pi}{q} +\frac\phi2\right)\right]^{(-1)^{m+n+1}}
\times
\prod_{m=2q+1}^{2q+n-1}
\left[\cos^2\left(\frac{p m \pi}{q} +\frac\phi2\right)\right]^{(-1)^{m+n+1}}
\\=&
\prod_{m=1}^{2q}
\left[\cos^2\left(\frac{p m \pi}{q} +\frac\phi2\right)\right]^{(-1)^{m+n+1}}
\times
\prod_{m=1}^{n-1}
\left[\cos^2\left(\frac{p m \pi}{q} +\frac\phi2+ 2p \pi\right)\right]^{(-1)^{m+n+1}}
,
\end{align}
which gives
\begin{align}
{\alpha_{n+2q}}=&
\prod_{m=1}^{2q}
\left[\cos^2\left(\frac{p m \pi}{q} +\frac\phi2\right)\right]^{(-1)^{m+n+1}}
\times
{\alpha_{n}}
,
\end{align}
Hence, to demonstrate the periodicity of ${\alpha_{n}}$, one needs to demonstrate that
\begin{equation}
\prod_{m=1}^{2q}
\left[\cos^2\left(\frac{p m \pi}{q} +\frac\phi2\right)\right]^{(-1)^{m}}
=1
.
\end{equation}
or equivalently that
\begin{equation}
\left\vert
\prod_{m=1}^{q}
\cos\left(\frac{2m p \pi}{q} +\frac\phi2\right)
\right\vert
=
\left\vert
\prod_{m=1}^{q}
\cos\left(\frac{2m p \pi}{q} +\frac\phi2+\frac{\pi}q\right)
\right\vert
.
\end{equation}

Let us consider the B\'ezout's identity $ax + by = 1$, which has integer solutions if $x,y$ are coprimes.
Thus, for $x=2$ and $y=q$, which are coprimes ($q$ is odd by assumption), one gets that $2l+kq=1$ has the integer solution $l,k$.
Moreover, the integer $k$ is odd; otherwise, $l+kq/2$ is an integer, which cannot be since $l+kq/2=1/2$.
Hence
\begin{gather}
\prod_{m=1}^{q}
\cos\left(\frac{2m p \pi}{q} +\frac\phi2+\frac{\pi}q\right)
=
\prod_{m=1}^{q}
\cos\left(\frac{2m p \pi}{q} +\frac\phi2+\frac{2\pi l}q +k\pi\right)
=
(-1)^k
\prod_{m=1}^{q}
\cos\left(\frac{2(mp+l) \pi}{q} +\frac\phi2\right)
.
\end{gather}
Now, since the order of the factors in the product is irrelevant, one can rearrange them by redefining $m$ and obtain
\begin{gather}
\prod_{m=1}^{q}
\cos\left(\frac{2m p \pi}{q} +\frac\phi2+\frac{\pi}q\right)
=
-
\prod_{m=1}^{q}
\cos\left(\frac{2m \pi}{q} +\frac\phi2\right)
,
\end{gather}
which thus ends the proof.

Notice that this proof cannot be extended to the case where $q$ is even since, in that case, the equation $2l+kq=1$ has no integer solutions.
Hence, assuming $\omega=2\pi p/q$ with $p,q$ coprimes, $\alpha_n$ is periodic in the lattice index $n$ with period $2q$ when $q$ is odd.
Conversely, $\alpha_n$ is not, in general, periodic when $q$ is even.


\section{Approximate form of the metric for $p=1$ and large $q$}

\begin{figure*}[t]
 \centering
	\includegraphics[width=.4\textwidth]{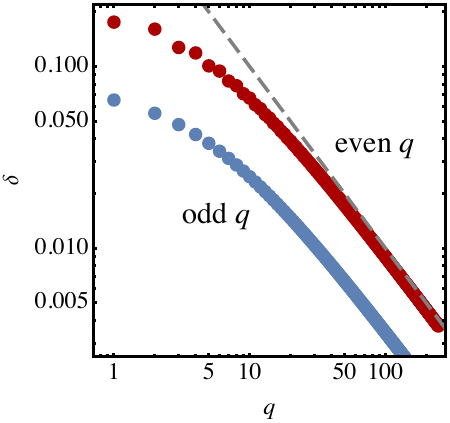}
	\caption{
The approximation error $\delta$ as defined in \cref{eq:metriclimiterror} corresponding with approximating the metric $\alpha_n$ with \cref{eq:metriclimit} plotted on a log-log scale as a function of $q$ by taking $\phi=0$ when $q$ is odd, and $\phi=\pi/q$ when $q$ is even in \cref{eq:specialmetric}.
The error scales polynomially as $\propto1/q$ (dashed line).
}
\label{fig:scaling}
\end{figure*}

For large values of $q$, by numerical investigation, I found that for the simple case $p=1$, and by taking $\phi=0$ in \cref{eq:specialmetric} when $q$ is odd, and $\phi=\pi/q$ when $q$ is even, the metric can be approximated by \cref{eq:metriclimit}, and that the approximation error scales as
\begin{equation}
\delta=\max_n\left(\left\vert\left\vert\cos(\frac{n\pi}q - \frac{\pi}{2q}+\frac{\phi}2)\right\vert
-\alpha_n\right\vert\right)
\propto\frac1q,
\label{eq:metriclimiterror}
\end{equation}
as shown numerically in \cref{fig:scaling}.

\section{Explicit form of the metric for small $p$ and $q$}

The metric $\alpha_n$ in \cref{eq:specialmetric} can be calculated explicitly for small values of $p$ and $q$.
For $p/q=1/2$ and $C=1$, one obtains
\begin{equation}
\begin{array}{rl}
 \alpha _1= & 1 \\
 \alpha _2= & \sin ^2\left(\frac{\phi}{2}\right) \\
 \alpha _3= & \cot ^2\left(\frac{\phi}{2}\right) \\
 \alpha _4= & \sin ^2\left(\frac{\phi}{2}\right) \tan ^2\left(\frac{\phi}{2}\right) \\
 \alpha _5= & \cot ^4\left(\frac{\phi}{2}\right) \\
 \alpha _6= & \sin ^2\left(\frac{\phi}{2}\right) \tan ^4\left(\frac{\phi}{2}\right) \\
 \alpha _7= & \cot ^6\left(\frac{\phi}{2}\right) \\
 \alpha _8= & \sin ^2\left(\frac{\phi}{2}\right) \tan ^6\left(\frac{\phi}{2}\right) \\
\ldots
\end{array}
\end{equation}
This metric is not periodic for arbitrary phases $\phi$.
For $\phi=\pi/2+m\pi$ with $m\in\mathbb{Z}$, the metric is periodic with $ \alpha _1= 1$, $\alpha _2=1/2$, $\alpha _{n+2}= \alpha _n$.

For $p/q=1/3$ and $C=1$, one obtains
\begin{equation}
\begin{array}{rl}
 \alpha _1= & 1 \\
 \alpha _2= & \sin ^2\left(\frac{\pi}{6}-\frac{\phi}{2}\right) \\
 \alpha _3= & \sin ^2\left(\frac{\phi}{2}+\frac{\pi}{6}\right) \csc ^2\left(\frac{\pi
}{6}-\frac{\phi}{2}\right) \\
 \alpha _4= & \sin ^2\left(\frac{\pi}{6}-\frac{\phi}{2}\right) \cos ^2\left(\frac{\phi
}{2}\right) \csc ^2\left(\frac{\phi}{2}+\frac{\pi}{6}\right) \\
 \alpha _5= & \sin ^2\left(\frac{\phi}{2}+\frac{\pi}{6}\right) \sec ^2\left(\frac{\phi
}{2}\right) \\
 \alpha _6= & \cos ^2\left(\frac{\phi}{2}\right)
\end{array}
\end{equation}
For $p/q=2/3$ and $C=1$, one obtains
\begin{equation}
\begin{array}{rl}
 \alpha _1= & 1 \\
 \alpha _2= & \sin ^2\left(\frac{\phi}{2}+\frac{\pi}{6}\right) \\
 \alpha _3= & \sin ^2\left(\frac{\pi}{6}-\frac{\phi}{2}\right) \csc ^2\left(\frac{\phi
}{2}+\frac{\pi}{6}\right) \\
 \alpha _4= & \frac{1}{16} \left(2 \sin \left(\phi +\frac{\pi}{6}\right)+1\right)^2 \csc
 ^2\left(\frac{\pi}{6}-\frac{\phi}{2}\right) \\
 \alpha _5= & \sin ^2\left(\frac{\pi}{6}-\frac{\phi}{2}\right) \sec ^2\left(\frac{\phi
}{2}\right) \\
 \alpha _6= & \cos ^2\left(\frac{\phi}{2}\right)
\end{array}
\end{equation}
In these cases $p/q=1/3,2/3$, the metric is periodic with $\alpha_{n+6}=\alpha_n$.

For $p/q=1/4$ and $C=1$, one obtains
\begin{equation}
\begin{array}{rl}
 \alpha _1= & 1 \\
 \alpha _2= & \cos ^2\left(\frac{\phi}{2}+\frac{\pi}{4}\right) \\
 \alpha _3= & \sin ^2\left(\frac{\phi}{2}\right) \csc ^2\left(\frac{\pi}{4}-\frac{\phi}{2}\right) \\
 \alpha _4= & \sin ^2\left(\frac{\pi}{4}-\frac{\phi}{2}\right) \sin ^2\left(\frac{\phi}{2}+\frac{\pi}{4}\right) \csc ^2\left(\frac{\phi}{2}\right) \\
 \alpha _5= & \tan ^2(\phi ) \\
 \alpha _6= & \sin ^4\left(\frac{\pi}{4}-\frac{\phi}{2}\right) \sin ^2\left(\frac{\phi}{2}+\frac{\pi}{4}\right) \csc ^2\left(\frac{\phi}{2}\right) \sec ^2\left(\frac{\phi}{2}\right) \\
 \alpha _7= & \sin ^4\left(\frac{\phi}{2}\right) \cos ^2\left(\frac{\phi}{2}\right) \csc ^4\left(\frac{\pi}{4}-\frac{\phi}{2}\right) \csc ^2\left(\frac{\phi}{2}+\frac{\pi}{4}\right) \\
 \alpha _8= & \sin ^4\left(\frac{\pi}{4}-\frac{\phi}{2}\right) \sin ^4\left(\frac{\phi}{2}+\frac{\pi}{4}\right) \csc ^4\left(\frac{\phi}{2}\right) \sec ^2\left(\frac{\phi}{2}\right) \\
 \alpha _9= & \tan ^4(\phi ) \\
 \alpha _{10}= & \sin ^6\left(\frac{\pi}{4}-\frac{\phi}{2}\right) \sin ^4\left(\frac{\phi}{2}+\frac{\pi}{4}\right) \csc ^4\left(\frac{\phi}{2}\right) \sec ^4\left(\frac{\phi}{2}\right) \\
 \alpha _{11}= & \sin ^6\left(\frac{\phi}{2}\right) \cos ^4\left(\frac{\phi}{2}\right) \csc ^6\left(\frac{\pi}{4}-\frac{\phi}{2}\right) \csc ^4\left(\frac{\phi}{2}+\frac{\pi}{4}\right) \\
 \alpha _{12}= & \sin ^6\left(\frac{\pi}{4}-\frac{\phi}{2}\right) \sin ^6\left(\frac{\phi}{2}+\frac{\pi}{4}\right) \csc ^6\left(\frac{\phi}{2}\right) \sec ^4\left(\frac{\phi}{2}\right) \\
 \alpha _{13}= & \tan ^6(\phi ) \\
 \alpha _{14}= & \sin ^8\left(\frac{\pi}{4}-\frac{\phi}{2}\right) \sin ^6\left(\frac{\phi}{2}+\frac{\pi}{4}\right) \csc ^6\left(\frac{\phi}{2}\right) \sec ^6\left(\frac{\phi}{2}\right) \\
 \alpha _{15}= & \sin ^8\left(\frac{\phi}{2}\right) \cos ^6\left(\frac{\phi}{2}\right) \csc ^8\left(\frac{\pi}{4}-\frac{\phi}{2}\right) \csc ^6\left(\frac{\phi}{2}+\frac{\pi}{4}\right) \\
 \alpha _{16}= & \sin ^8\left(\frac{\pi}{4}-\frac{\phi}{2}\right) \sin ^8\left(\frac{\phi}{2}+\frac{\pi}{4}\right) \csc ^8\left(\frac{\phi}{2}\right) \sec ^6\left(\frac{\phi}{2}\right) \\
 \ldots
\end{array}
\end{equation}
This metric is not periodic for arbitrary phases $\phi$.
For $\phi=\pi/4+m\pi$ with $m\in\mathbb{Z}$, the metric is periodic with $ \alpha _1= 1$, $\alpha _2= \sin ^2\left(\frac{\pi}{8}\right)$, $\alpha _3=1$, $\alpha _4= \cos ^2\left(\frac{\pi}{8}\right)$, $\alpha _{n+4}= \alpha _n$.
For $p/q=3/4$ and $C=1$, one obtains
\begin{equation}
\begin{array}{rl}
 \alpha _1= & 1 \\
 \alpha _2= & \sin ^2\left(\frac{1}{4} (2 \phi +\pi )\right) \\
 \alpha _3= & \frac{1-\cos (\phi )}{\sin (\phi )+1} \\
 \alpha _4= & \frac{1}{4} \cos ^2(\phi ) \csc ^2\left(\frac{\phi}{2}\right) \\
 \alpha _5= & \tan ^2(\phi ) \\
 \alpha _6= & \frac{1}{2} \sin (\phi ) (\csc (\phi )-1) (\csc (\phi )+1)^2 \\
 \alpha _7= & \sin ^2\left(\frac{\phi}{2}\right) \tan ^2(\phi ) \csc ^2\left(\frac{1}{4} (2 \phi +\pi )\right) \\
 \alpha _8= & \frac{1}{4} \cos ^2(\phi ) \cot ^2(\phi ) \csc ^2\left(\frac{\phi}{2}\right) \\
 \alpha _9= & \tan ^4(\phi ) \\
 \alpha _{10}= & \frac{1}{2} \sin (\phi ) (\csc (\phi )-1)^2 (\csc (\phi )+1)^3 \\
 \alpha _{11}= & \sin ^2\left(\frac{\phi}{2}\right) \tan ^4(\phi ) \csc ^2\left(\frac{1}{4} (2 \phi +\pi )\right) \\
 \alpha _{12}= & \frac{1}{64} \cos ^6(\phi ) \csc ^6\left(\frac{\phi}{2}\right) \sec ^4\left(\frac{\phi}{2}\right) \\
 \alpha _{13}= & \tan ^6(\phi ) \\
 \alpha _{14}= & \frac{1}{2} \sin (\phi ) (\csc (\phi )-1)^3 (\csc (\phi )+1)^4 \\
 \alpha _{15}= & \sin ^2\left(\frac{\phi}{2}\right) \tan ^6(\phi ) \csc ^2\left(\frac{1}{4} (2 \phi +\pi )\right) \\
 \alpha _{16}= & \frac{1}{256} \cos ^8(\phi ) \csc ^8\left(\frac{\phi}{2}\right) \sec ^6\left(\frac{\phi}{2}\right) \\
 \ldots
\end{array}
\end{equation}
This metric is not periodic.

For $p/q=1/5$ and $C=1$, one obtains
\begin{equation}
\begin{array}{rl}
 \alpha _1= & 1 \\
 \alpha _2= & \cos ^2\left(\frac{\phi}{2}+\frac{\pi}{5}\right) \\
 \alpha _3= & \sin ^2\left(\frac{\pi}{10}-\frac{\phi}{2}\right) \sec ^2\left(\frac{\phi}{2}+\frac{\pi}{5}\right) \\
 \alpha _4= & \sin ^2\left(\frac{\phi}{2}+\frac{\pi}{10}\right) \cos ^2\left(\frac{\phi}{2}+\frac{\pi}{5}\right) \csc ^2\left(\frac{\pi}{10}-\frac{\phi}{2}\right) \\
 \alpha _5= & \frac{\left(-4 \cos \left(\phi +\frac{\pi}{5}\right)+\sqrt{5}-1\right)^2}{\left(-4 \cos \left(\frac{\pi}{5}-\phi \right)+\sqrt{5}-1\right)^2} \\
 \alpha _6= & \sin ^2\left(\frac{\phi}{2}+\frac{\pi}{10}\right) \cos ^2\left(\frac{\phi}{2}+\frac{\pi}{5}\right) \cos ^2\left(\frac{\phi}{2}\right) \csc ^2\left(\frac{\pi}{10}-\frac{\phi}{2}\right) \sec ^2\left(\frac{\pi}{5}-\frac{\phi}{2}\right) \\
 \alpha _7= & \sin ^2\left(\frac{\pi}{10}-\frac{\phi}{2}\right) \cos ^2\left(\frac{\pi}{5}-\frac{\phi}{2}\right) \csc ^2\left(\frac{\phi}{2}+\frac{\pi}{10}\right) \sec ^2\left(\frac{\phi}{2}\right) \\
 \alpha _8= & \sin ^2\left(\frac{\phi}{2}+\frac{\pi}{10}\right) \cos ^2\left(\frac{\phi}{2}\right) \sec ^2\left(\frac{\pi}{5}-\frac{\phi}{2}\right) \\
 \alpha _9= & \cos ^2\left(\frac{\pi}{5}-\frac{\phi}{2}\right) \sec ^2\left(\frac{\phi}{2}\right) \\
 \alpha _{10}= & \cos ^2\left(\frac{\phi}{2}\right)
 \end{array}
\end{equation}
For $p/q=2/5$ and $C=1$, one obtains
\begin{equation}
\begin{array}{rl}
 \alpha _1= & 1 \\
 \alpha _2= & \sin ^2\left(\frac{\pi}{10}-\frac{\phi}{2}\right) \\
 \alpha _3= & \cos ^2\left(\frac{\pi}{5}-\frac{\phi}{2}\right) \csc ^2\left(\frac{\pi}{10}-\frac{\phi}{2}\right) \\
 \alpha _4= & \frac{1}{64} \left(-4 \sin \left(\phi +\frac{\pi}{10}\right)+\sqrt{5}+1\right)^2 \sec ^2\left(\frac{\pi}{5}-\frac{\phi}{2}\right) \\
 \alpha _5= & \frac{\left(-4 \sin \left(\frac{\pi}{10}-\phi \right)+\sqrt{5}+1\right)^2}{\left(-4 \sin \left(\phi +\frac{\pi}{10}\right)+\sqrt{5}+1\right)^2} \\
 \alpha _6= & \sin ^2\left(\frac{\pi}{10}-\frac{\phi}{2}\right) \cos ^2\left(\frac{\phi}{2}+\frac{\pi}{5}\right) \cos ^2\left(\frac{\phi}{2}\right) \csc ^2\left(\frac{\phi}{2}+\frac{\pi}{10}\right) \sec ^2\left(\frac{\pi}{5}-\frac{\phi}{2}\right) \\
 \alpha _7= & \sin ^2\left(\frac{\phi}{2}+\frac{\pi}{10}\right) \cos ^2\left(\frac{\pi}{5}-\frac{\phi}{2}\right) \sec ^2\left(\frac{\phi}{2}+\frac{\pi}{5}\right) \sec ^2\left(\frac{\phi}{2}\right) \\
 \alpha _8= & \cos ^2\left(\frac{\phi}{2}+\frac{\pi}{5}\right) \cos ^2\left(\frac{\phi}{2}\right) \csc ^2\left(\frac{\phi}{2}+\frac{\pi}{10}\right) \\
 \alpha _9= & \sin ^2\left(\frac{\phi}{2}+\frac{\pi}{10}\right) \sec ^2\left(\frac{\phi}{2}\right) \\
 \alpha _{10}= & \cos ^2\left(\frac{\phi}{2}\right)
 \end{array}
\end{equation}
For $p/q=3/5$ and $C=1$, one obtains
\begin{equation}
\begin{array}{rl}
 \alpha _1= & 1 \\
 \alpha _2= & \sin ^2\left(\frac{\phi}{2}+\frac{\pi}{10}\right) \\
 \alpha _3= & \cos ^2\left(\frac{\phi}{2}+\frac{\pi}{5}\right) \csc ^2\left(\frac{\phi}{2}+\frac{\pi}{10}\right) \\
 \alpha _4= & \sin ^2\left(\frac{\phi}{2}+\frac{\pi}{10}\right) \cos ^2\left(\frac{\pi}{5}-\frac{\phi}{2}\right) \sec ^2\left(\frac{\phi}{2}+\frac{\pi}{5}\right) \\
 \alpha _5= & \frac{\left(-4 \sin \left(\phi +\frac{\pi}{10}\right)+\sqrt{5}+1\right)^2}{\left(-4 \sin \left(\frac{\pi}{10}-\phi \right)+\sqrt{5}+1\right)^2} \\
 \alpha _6= & \sin ^2\left(\frac{\phi}{2}+\frac{\pi}{10}\right) \cos ^2\left(\frac{\pi}{5}-\frac{\phi}{2}\right) \cos ^2\left(\frac{\phi}{2}\right) \csc ^2\left(\frac{\pi}{10}-\frac{\phi}{2}\right) \sec ^2\left(\frac{\phi}{2}+\frac{\pi}{5}\right) \\
 \alpha _7= & \frac{\left(-4 \sin \left(\phi +\frac{\pi}{10}\right)+\sqrt{5}+1\right)^2}{\left(4 \cos \left(\frac{\pi}{5}-\phi \right)+\sqrt{5}+1\right)^2} \\
 \alpha _8= & \cos ^2\left(\frac{\pi}{5}-\frac{\phi}{2}\right) \cos ^2\left(\frac{\phi}{2}\right) \csc ^2\left(\frac{\pi}{10}-\frac{\phi}{2}\right) \\
 \alpha _9= & \sin ^2\left(\frac{\pi}{10}-\frac{\phi}{2}\right) \sec ^2\left(\frac{\phi}{2}\right) \\
 \alpha _{10}= & \cos ^2\left(\frac{\phi}{2}\right)
 \end{array}
\end{equation}
For $p/q=4/5$ and $C=1$, one obtains
\begin{equation}
\begin{array}{rl}
 \alpha _1= & 1 \\
 \alpha _2= & \cos ^2\left(\frac{\pi}{5}-\frac{\phi}{2}\right) \\
 \alpha _3= & \sin ^2\left(\frac{\phi}{2}+\frac{\pi}{10}\right) \sec ^2\left(\frac{\pi}{5}-\frac{\phi}{2}\right) \\
 \alpha _4= & \frac{1}{64} \left(-4 \cos \left(\phi +\frac{\pi}{5}\right)+\sqrt{5}-1\right)^2 \csc ^2\left(\frac{\phi}{2}+\frac{\pi}{10}\right) \\
 \alpha _5= & \frac{\left(-4 \cos \left(\frac{\pi}{5}-\phi \right)+\sqrt{5}-1\right)^2}{\left(-4 \cos \left(\phi +\frac{\pi}{5}\right)+\sqrt{5}-1\right)^2} \\
 \alpha _6= & \sin ^2\left(\frac{\pi}{10}-\frac{\phi}{2}\right) \cos ^2\left(\frac{\pi}{5}-\frac{\phi}{2}\right) \cos ^2\left(\frac{\phi}{2}\right) \csc ^2\left(\frac{\phi}{2}+\frac{\pi}{10}\right) \sec ^2\left(\frac{\phi}{2}+\frac{\pi}{5}\right) \\
 \alpha _7= & \sin ^2\left(\frac{\phi}{2}+\frac{\pi}{10}\right) \cos ^2\left(\frac{\phi}{2}+\frac{\pi}{5}\right) \csc ^2\left(\frac{\pi}{10}-\frac{\phi}{2}\right) \sec ^2\left(\frac{\phi}{2}\right) \\
 \alpha _8= & \sin ^2\left(\frac{\pi}{10}-\frac{\phi}{2}\right) \cos ^2\left(\frac{\phi}{2}\right) \sec ^2\left(\frac{\phi}{2}+\frac{\pi}{5}\right) \\
 \alpha _9= & \cos ^2\left(\frac{\phi}{2}+\frac{\pi}{5}\right) \sec ^2\left(\frac{\phi}{2}\right) \\
 \alpha _{10}= & \cos ^2\left(\frac{\phi}{2}\right) \\
\end{array}
\end{equation}
In these cases, the metric is periodic with $\alpha_{n+10}=\alpha_n$.


\end{document}